# Adaptive Feed Rate Policies for Spiral Drilling Using Markov Decision Process

Yedige Tlegenov, Wong Yoke San, Hong Geok Soon
*Department of Mechanical Engineering*
*National University of Singapore*
*9 Engineering Drive 1, Singapore, 117576, Singapore*
yedige@nus.edu.sg, mpewys@nus.edu.sg, mpehgs@nus.edu.sg,

*Abstract*-**In this study, the feed rate optimization model based on a Markov Decision Process (MDP) was introduced for spiral drilling process. Firstly, the experimental data on spiral drilling was taken from literature for different axial force parameters and with various feed rate decisions made, having the length of a hole being drilled as a reward. Proposed optimization model was computed using value iteration method. Secondly, the results of computations were displayed for optimal decision to be made on each state. Proposed decisions for an optimal feed rate could be utilized in order to improve the efficiency of spiral drilling process in terms of cost and time.**

## I. Introduction

Spiral drilling is a process that produces straight holes with the ratio of hole depth to hole diameter less than ten. Nowadays spiral drilling has lots of applications in automotive, aeronautics, hydraulic, petrochemical and oil & gas industries. As an illustration, automotive industries are performing drilling process in order to manufacture various types of engine parts. Perforating a number of holes in an engine parts requires precise choice of the feed rates in order to maintain stability of the drill bit. Depending upon different feed rates the axial force is subjected to changes, which itself tends to the variation of the cutting speed. It is said that the cutting speed has an influence on the length of the hole being drilled [1]. In other words, even slight changes in feed rates cause significant changes on the length of the drilled hole.

This paper aims to present an optimal feed rate choice policy for spiral drilling system using MDP [2, 3]. The data required for the MDP model include axial force values and number of feed rates corresponding to each this value, length of the hole drilled for each of the feed rate value, corresponding transition probabilities [1]. Depending upon certain axial force value a feed rate is to be chosen for drilling hole with maximum possible length.

## II. Case Study on Spiral Drilling

Spiral Drilling of holes in an automotive engine parts require getting sufficient quality of holes in terms of surface finish and straightness. The spiral drilling system is required to be time consumable to increase productivity of the work. To meet these requirements for quality and productivity the system is proposed to have adaptive feed rate choice policy using MDP to identify any significant change in the tool parameters and proceed with the most optimal parameter. At the time when changes in axial force identified, the system is set to make immediate and optimal action for the feed rate. The data required for the MDP model of transformers was taken from experimental results for spiral drilling using a number of different tryouts [1]. For each set of the axial force values there are five sets of different feed rates along with final length of the hole being drilled for each of these chosen parameters. Axial force values are set as conditions for given problem, so there are ten conditions for description of the axial force in total: $F_1, F_2, F_3, F_4, F_5, F_6, F_7, F_8, F_9, F_{10}$. For each of the axial force parameter there are five different feed rates, which are considered as set of actions: $S_1^1, S_2^1, S_3^1, S_4^1, S_5^1, ..., S_1^{10}, S_2^{10}, S_3^{10}, S_4^{10}, S_5^{10}$. For each of the parameter chosen for feed rate in particular axial force value there is a result which expressed as a length of the drilled hole. This result is considered as a reward for given problem: $L_1^1, L_2^1, L_3^1, L_4^1, L_5^1, ..., L_1^{10}, L_2^{10}, L_3^{10}, L_4^{10}, L_5^{10}$. The data for MDP model is presented in Table 1.

TABLE I

DATA FOR MDP MODEL OF DRILLING DEEP HOLES WITH A SPIRAL DRILL

| State I, Axial force rates per drilled hole - F, Newton | | Decision k, | Feed rate - S, mm/rev | Transition probabilities | | | | | | | | | | Reward q - Length of hole=L, x10$^{-2}$ mm |
|---|---|---|---|---|---|---|---|---|---|---|---|---|---|---|
| | | | | P1 | P2 | P3 | P4 | P5 | P6 | P7 | P8 | P9 | P10 | |
| 1 | 50.92 | 1 | 0.0522 | 0.75 | 0.15 | 0.1 | 0 | 0 | 0 | 0 | 0 | 0 | 0 | 7161.82 |
| | | 2 | 0.0582 | 0.65 | 0.2 | 0.15 | 0 | 0 | 0 | 0 | 0 | 0 | 0 | 7430.09 |
| | | 3 | 0.0658 | 0.5 | 0.5 | 0 | 0 | 0 | 0 | 0 | 0 | 0 | 0 | 5009.24 |
| | | 4 | 0.0742 | 0.2 | 0.65 | 0.15 | 0 | 0 | 0 | 0 | 0 | 0 | 0 | 2056.31 |
| | | 5 | 0.0831 | 0.15 | 0.75 | 0.1 | 0 | 0 | 0 | 0 | 0 | 0 | 0 | 498.15 |
| 2 | 83.75 | 1 | 0.0832 | 0 | 0.75 | 0.15 | 0.1 | 0 | 0 | 0 | 0 | 0 | 0 | 8278.23 |
| | | 2 | 0.0931 | 0 | 0.65 | 0.2 | 0.15 | 0 | 0 | 0 | 0 | 0 | 0 | 8500.39 |
| | | 3 | 0.1053 | 0 | 0.5 | 0.5 | 0 | 0 | 0 | 0 | 0 | 0 | 0 | 6036.86 |
| | | 4 | 0.1185 | 0 | 0.2 | 0.65 | 0.15 | 0 | 0 | 0 | 0 | 0 | 0 | 2827.92 |
| | | 5 | 0.1324 | 0 | 0.15 | 0.75 | 0.1 | 0 | 0 | 0 | 0 | 0 | 0 | 854.43 |
| 3 | 112.12 | 1 | 0.1029 | 0 | 0 | 0.75 | 0.15 | 0.1 | 0 | 0 | 0 | 0 | 0 | 7831.99 |
| | | 2 | 0.1162 | 0 | 0 | 0.65 | 0.2 | 0.15 | 0 | 0 | 0 | 0 | 0 | 9450.55 |
| | | 3 | 0.1341 | 0 | 0 | 0.5 | 0.5 | 0 | 0 | 0 | 0 | 0 | 0 | 7471.69 |
| | | 4 | 0.154 | 0 | 0 | 0.2 | 0.65 | 0.15 | 0 | 0 | 0 | 0 | 0 | 3568.89 |
| | | 5 | 0.1749 | 0 | 0 | 0.15 | 0.75 | 0.1 | 0 | 0 | 0 | 0 | 0 | 995.89 |
| 4 | 147.36 | 1 | 0.1254 | 0 | 0 | 0 | 0.75 | 0.15 | 0.1 | 0 | 0 | 0 | 0 | 9228.49 |
| | | 2 | 0.1481 | 0 | 0 | 0 | 0.65 | 0.2 | 0.15 | 0 | 0 | 0 | 0 | 9141.06 |
| | | 3 | 0.1753 | 0 | 0 | 0 | 0.5 | 0.5 | 0 | 0 | 0 | 0 | 0 | 4731.06 |
| | | 4 | 0.2043 | 0 | 0 | 0 | 0.2 | 0.65 | 0.15 | 0 | 0 | 0 | 0 | 1192.48 |
| | | 5 | 0.2341 | 0 | 0 | 0 | 0.15 | 0.75 | 0.1 | 0 | 0 | 0 | 0 | 142.72 |
| 5 | 159.54 | 1 | 0.1316 | 0 | 0 | 0 | 0 | 0.75 | 0.15 | 0.1 | 0 | 0 | 0 | 8061.25 |
| | | 2 | 0.1513 | 0 | 0 | 0 | 0 | 0.65 | 0.2 | 0.15 | 0 | 0 | 0 | 10198.16 |
| | | 3 | 0.1781 | 0 | 0 | 0 | 0 | 0.5 | 0.5 | 0 | 0 | 0 | 0 | 7858.75 |
| | | 4 | 0.2077 | 0 | 0 | 0 | 0 | 0.2 | 0.65 | 0.15 | 0 | 0 | 0 | 3344.22 |
| | | 5 | 0.2387 | 0 | 0 | 0 | 0 | 0.15 | 0.75 | 0.1 | 0 | 0 | 0 | 758.47 |
| 6 | 187.06 | 1 | 0.1461 | 0 | 0 | 0 | 0 | 0 | 0.75 | 0.15 | 0.1 | 0 | 0 | 8909.61 |
| | | 2 | 0.1729 | 0 | 0 | 0 | 0 | 0 | 0.65 | 0.2 | 0.15 | 0 | 0 | 10270.01 |
| | | 3 | 0.2067 | 0 | 0 | 0 | 0 | 0 | 0.5 | 0.5 | 0 | 0 | 0 | 6290.49 |
| | | 4 | 0.2431 | 0 | 0 | 0 | 0 | 0 | 0.2 | 0.65 | 0.15 | 0 | 0 | 1871.15 |

|   |   | 5 | 0.2807 | 0 | 0 | 0 | 0 | 0 | 0.15 | 0.75 | 0.1 | 0 | 0 | 262.34 |
|---|---|---|--------|---|---|---|---|---|------|------|-----|---|---|--------|
| 7 | 203.92 | 1 | 0.1602 | 0 | 0 | 0 | 0 | 0 | 0 | 0.75 | 0.15 | 0.1 | 0 | 9622.59 |
|   |   | 2 | 0.1863 | 0 | 0 | 0 | 0 | 0 | 0 | 0.65 | 0.2 | 0.15 | 0 | 10449.93 |
|   |   | 3 | 0.2177 | 0 | 0 | 0 | 0 | 0 | 0 | 0.5 | 0.5 | 0 | 0 | 7090.61 |
|   |   | 4 | 0.2512 | 0 | 0 | 0 | 0 | 0 | 0 | 0.2 | 0.65 | 0.15 | 0 | 2841.58 |
|   |   | 5 | 0.2858 | 0 | 0 | 0 | 0 | 0 | 0 | 0.15 | 0.75 | 0.1 | 0 | 658.41 |
| 8 | 228.71 | 1 | 0.1738 | 0 | 0 | 0 | 0 | 0 | 0 | 0 | 0.75 | 0.15 | 0.1 | 10206.75 |
|   |   | 2 | 0.2053 | 0 | 0 | 0 | 0 | 0 | 0 | 0 | 0.65 | 0.2 | 0.15 | 10166.77 |
|   |   | 3 | 0.2418 | 0 | 0 | 0 | 0 | 0 | 0 | 0 | 0.5 | 0.5 | 0 | 5848.13 |
|   |   | 4 | 0.2804 | 0 | 0 | 0 | 0 | 0 | 0 | 0 | 0.2 | 0.65 | 0.15 | 1845.82 |
|   |   | 5 | 0.3199 | 0 | 0 | 0 | 0 | 0 | 0 | 0 | 0.15 | 0.75 | 0.1 | 313.76 |
| 9 | 240.35 | 1 | 0.173 | 0 | 0 | 0 | 0 | 0 | 0 | 0 | 0 | 0.75 | 0.25 | 9206.44 |
|   |   | 2 | 0.2046 | 0 | 0 | 0 | 0 | 0 | 0 | 0 | 0 | 0.65 | 0.35 | 10927.64 |
|   |   | 3 | 0.2442 | 0 | 0 | 0 | 0 | 0 | 0 | 0 | 0 | 0.5 | 0.5 | 7327.75 |
|   |   | 4 | 0.2869 | 0 | 0 | 0 | 0 | 0 | 0 | 0 | 0 | 0.35 | 0.65 | 2556.97 |
|   |   | 5 | 0.3308 | 0 | 0 | 0 | 0 | 0 | 0 | 0 | 0 | 0.25 | 0.75 | 451.9 |
| 10 | 262.94 | 1 | 0.1835 | 0 | 0 | 0 | 0 | 0 | 0 | 0 | 0 | 0.2 | 0.8 | 9701.99 |
|   |   | 2 | 0.2203 | 0 | 0 | 0 | 0 | 0 | 0 | 0 | 0 | 0.15 | 0.85 | 10842.65 |
|   |   | 3 | 0.2648 | 0 | 0 | 0 | 0 | 0 | 0 | 0 | 0 | 0.1 | 0.9 | 6355.54 |
|   |   | 4 | 0.3122 | 0 | 0 | 0 | 0 | 0 | 0 | 0 | 0 | 0.05 | 0.95 | 1810.01 |
|   |   | 5 | 0.3608 | 0 | 0 | 0 | 0 | 0 | 0 | 0 | 0 | 0 | 1 | 244.43 |

The MDP model based on data shown in Table 1 is solved using backward induction algorithm. This method is applied in order to determine which decision to make in every state in each drilled hole of a ten holes to be drilled planning horizon so that the vector of expected total rewards is maximized. Optimal policy for a finite horizon is defined as which maximizes the vector of expected total rewards received until the end of the horizon. An optimal policy can be found by utilizing a value iteration method.

For the present case study the value iteration equations are:

$$v_i(10) = 0, \text{ for } i = 1, 2, \ldots, 10$$

$$v_i(n) = \max_k \left[ q_i^k + \sum_{j=1}^{10} p_{ij}^k v_j(n+1) \right]$$

$$= \max_k [q_i^k + p_{i1}^k v_1(n+1) + p_{i2}^k v_2(n+1) + p_{i3}^k v_3(n+1) + p_{i4}^k v_4(n+1)$$
$$+ p_{i5}^k v_5(n+1) + p_{i6}^k v_6(n+1) + p_{i7}^k v_7(n+1) + p_{i8}^k v_8(n+1) + p_{i9}^k v_9(n+1)$$
$$+ p_{i10}^k v_{10}(n+1)]$$

for $n = 0, 1, \ldots, 9$, and $i = 1, 2, \ldots, 10$. Firstly, following values are specified for all states at the end of the hole number 10:

$$v_1(10) = v_2(10) = v_3(10) = v_4(10) = v_5(10) = v_6(10) = v_7(10) = v_8(10) = v_9(10) = v_{10}(10)$$
$$= 0$$

Since the value iteration is a form of dynamic programming, the calculations for each epoch is displayed in a tabular format.

The calculations for 9th hole are shown in Table 2.

$$v_i(9) = \max_k [q_i^k + p_{i1}^k v_1(10) + p_{i2}^k v_2(10) + p_{i3}^k v_3(10) + p_{i4}^k v_4(10) + p_{i5}^k v_5(10) + p_{i6}^k v_6(10)$$
$$+ p_{i7}^k v_7(10) + p_{i8}^k v_8(10) + p_{i9}^k v_9(10) + p_{i10}^k v_{10}(10)] = \max_k [q_i^k]$$

for $i = 1, 2, \ldots, 10$. At the end of hole 9, where n=9, the optimal decision is to select the maximum reward in each state. Consequently, the decision vector is computed as d(9)=[2 2 2 1 2 2 2 1 2 2]$^T$.

TABLE II
DATA VALUE ITERATION FOR N=9

| State | $q_i^1$ | $q_i^2$ | $q_i^3$ | $q_i^4$ | $q_i^5$ | Expected total reward, x10$^{-2}$ mm | Decision |
|---|---|---|---|---|---|---|---|
| i | k=1 | k=2 | k=3 | k=4 | k=5 | $v_i(9) = \max_k[q_i^1, q_i^2, q_i^3, q_i^4, q_i^5]$, for $i = 1, 2, \ldots, 10$ | k |
| 1 | 7161.82 | 7430.09 | 5009.24 | 2056.31 | 498.15 | $v_1(9) = \max_k[7161.82, 7430.09, 5009.24, 2056.31, 498.15] = 7430.09$ | 2 |
| 2 | 8278.23 | 8500.39 | 6036.86 | 2827.92 | 854.43 | $v_2(9) = \max_k[8278.23, 8500.39, 6036.86, 2827.92, 854.43] = 8500.39$ | 2 |
| 3 | 7831.99 | 9450.55 | 7471.69 | 3568.89 | 995.89 | $v_3(9) = \max_k[7831.99, 9450.55, 7471.69, 3568.89, 995.89] = 9450.55$ | 2 |
| 4 | 9228.49 | 9141.06 | 4731.06 | 1192.48 | 142.72 | $v_4(9) = \max_k[9228.49, 9141.06, 4731.06, 1192.48, 142.72] = 9228.49$ | 1 |
| 5 | 8061.25 | 10198.16 | 7858.75 | 3344.22 | 758.47 | $v_5(9) = \max_k[8061.25, 10198.16, 7858.75, 3344.22, 758.47] = 10198.16$ | 2 |
| 6 | 8909.61 | 10270.01 | 6290.49 | 1871.15 | 262.34 | $v_6(9) = \max_k[8909.61, 10270.01, 6290.49, 1871.15, 262.34] = 10270.01$ | 2 |
| 7 | 9622.59 | 10449.93 | 7090.61 | 2841.58 | 658.41 | $v_7(9) = \max_k[9622.59, 10449.93, 7090.61, 2841.58, 658.41] = 10449.93$ | 2 |
| 8 | 10206.75 | 10166.77 | 5848.13 | 1845.82 | 313.76 | $v_8(9) = \max_k[10206.75, 10166.77, 5848.13, 1845.82, 313.76] = 10206.75$ | 1 |
| 9 | 9206.44 | 10927.64 | 7327.75 | 2556.97 | 451.9 | $v_9(9) = \max_k[9206.44, 10927.64, 7327.75, 2556.97, 451.9] = 10927.64$ | 2 |
| 10 | 9701.99 | 10842.65 | 6355.54 | 1810.01 | 244.43 | $v_{10}(9) = \max_k[9701.99, 10842.65, 6255.54, 244.43] = 10842.65$ | 2 |

The calculations for 8th hole are shown in Table 3.

$$v_i(8) = \max_k [q_i^k + p_{i1}^k v_1(9) + p_{i2}^k v_2(9) + p_{i3}^k v_3(9) + p_{i4}^k v_4(9) + p_{i5}^k v_5(9) + p_{i6}^k v_6(9) + p_{i7}^k v_7(9)$$
$$+ p_{i8}^k v_8(9) + p_{i9}^k v_9(9) + p_{i10}^k v_{10}(9)]$$

$$v_i(8) = \max_k [q_i^k + p_{i1}^k(7430.09) + p_{i2}^k(5800.39) + p_{i3}^k(9450.55) + p_{i4}^k(9228.49) + p_{i5}^k(10198.16)$$
$$+ p_{i6}^k(10270.01) + p_{i7}^k(10449.93) + p_{i8}^k(10206.75) + p_{i9}^k v_9(10927.64)$$
$$+ p_{i,10}^k(10842.65)]$$

At the end of hole 8, where n=8, the optimal decision is to select the second alternative in each state. Consequently, the decision vector is computed as d(8)=[2 2 2 2 2 2 2 2 2]$^T$.

TABLE III
DATA VALUE ITERATION FOR N=8

| State i | | Decision k | | $v_i(8)$ | Expected total reward, x10$^{-2}$mm | Decision |
|---|---|---|---|---|---|---|
| 1 | 50.92 | 1 | 0.0522 | 7161.82+0.75(7430.09)+0.15(8500.39)+0.1(9450.55)=14954.5 | 15377.3 | 2 |
| | | 2 | 0.0582 | 7430.09+0.65(7430.09)+0.2(8500.39)+0.15(9450.55)=15377.3 | | |
| | | 3 | 0.0658 | 5009.24+0.5(7430.09)+0.5(8500.39)=12974.5 | | |
| | | 4 | 0.0742 | 2056.31+0.2(7430.09)+0.65(8500.39)+0.15(9450.55)=10485.2 | | |
| | | 5 | 0.0831 | 498.15+0.15(7430.09)+0.75(8500.39)+0.1(9450.55)=8933.01 | | |
| 2 | 83.75 | 1 | 0.0832 | 8278.23+0.75(8500.39)+0.15(9450.55)+0.1(9228.49)=16994 | 17300 | 2 |
| | | 2 | 0.0931 | 8500.39+0.65(8500.39)+0.2(9450.55)+0.15(9228.49)=17300 | | |
| | | 3 | 0.1053 | 6036.86+0.5(8500.39)+0.5(9450.55)=15012.3 | | |
| | | 4 | 0.1185 | 2827.92+0.2(8500.39)+0.65(9450.55)+0.15(9228.49)=12055.1 | | |
| | | 5 | 0.1324 | 854.43+0.15(8500.39)+0.75(9450.55)+0.1(9228.49)=10140.3 | | |
| 3 | 112.12 | 1 | 0.1029 | 7831.99+0.75(9450.55)+0.15(9228.49)+0.1(10198.2)=17324 | 18968.8 | 2 |
| | | 2 | 0.1162 | 9450.55+0.65(9450.55)+0.2(9228.49)+0.15(10198.2)=18968.8 | | |
| | | 3 | 0.1341 | 7471.69+0.5(9450.55)+0.5(9228.49)=16811.2 | | |
| | | 4 | 0.154 | 3568.89+0.2(9450.55)+0.65(9228.49)+0.15(10198.2)=12987.2 | | |
| | | 5 | 0.1749 | 995.89+0.15(9450.55)+0.75(9228.49)+0.1(10198.2)=10354.7 | | |
| 4 | 147.36 | 1 | 0.1254 | 9228.49+0.75(9228.49)+0.15(10198.2)+0.1(10270.01)=18706.6 | 18719.7 | 2 |
| | | 2 | 0.1481 | 9141.06+0.65(9228.49)+0.2(10198.2)+0.15(10270.01)=18719.7 | | |
| | | 3 | 0.1753 | 4731.06+0.5(9228.49)+0.5(10198.2)=14444.4 | | |
| | | 4 | 0.2043 | 1192.48+0.2(9228.49)+0.65(10198.2)+0.15(10270.01)=11207.5 | | |
| | | 5 | 0.2341 | 142.72+0.15(9228.49)+0.75(10198.2)+0.1(10270.01)=10202.6 | | |
| 5 | 159.54 | 1 | 0.1316 | 8061.25+0.75(10198.2)+0.15(10270.01)+0.1(10449.9)=18295.4 | 20448.5 | 2 |
| | | 2 | 0.1513 | 10198.16+0.65(10198.2)+0.2(10270.01)+0.15(10449.9)=20448.5 | | |
| | | 3 | 0.1781 | 7858.75+0.5(10198.2)+0.5(10270.01)=18092.8 | | |
| | | 4 | 0.2077 | 3344.22+0.2(10198.2)+0.65(10270.01)+0.15(10449.9)=13626.8 | | |
| | | 5 | 0.2387 | 758.47+0.15(10198.2)+0.75(10270.01)+0.1(10449.9)=11035.7 | | |
| 6 | 187.06 | 1 | 0.1461 | 8909.61+0.75(10270.01)+0.15(10449.9)+0.1(10206.8)=19200.3 | 20566.5 | 2 |
| | | 2 | 0.1729 | 10270.01+0.65(10270.01)+0.2(10449.9)+0.15(10206.8)=20566.5 | | |
| | | 3 | 0.2067 | 6290.49+0.5(10270.01)+0.5(10449.9)=16650.5 | | |
| | | 4 | 0.2431 | 1871.15+0.2(10270.01)+0.65(10449.9)+0.15(10206.8)=12248.6 | | |
| | | 5 | 0.2807 | 262.34+0.15(10270.01)+0.75(10449.9)+0.1(10206.8)=10661 | | |
| 7 | 203.92 | 1 | 0.1602 | 9622.59+0.75(10449.9)+0.15(10206.8)+0.1(10927.6)=20083.8 | 20922.9 | 2 |

| | | 2 | 0.1863 | 10449.93+0.65(10449.9)+0.2(10206.8)+0.15(10927.6)=20922.9 | | |
| | | 3 | 0.2177 | 7090.61+0.5(10449.9)+0.5(10206.8)=17419 | | |
| | | 4 | 0.2512 | 2841.58+0.2(10449.9)+0.65(10206.8)+0.15(10927.6)=13205.1 | | |
| | | 5 | 0.2858 | 658.41+0.15(10449.9)+0.75(10206.8)+0.1(10927.6)=10973.7 | | |
| 8 | 228.71 | 1 | 0.1738 | 10206.75+0.75(10206.8)+0.15(10927.6)+0.1(10842.7)=20585.2 | 20613.1 | 2 |
| | | 2 | 0.2053 | 10166.77+0.65(10206.8)+0.2(10927.6)+0.15(10842.7)=20613.1 | | |
| | | 3 | 0.2418 | 5848.13+0.5(10206.8)+0.5(10927.6)=16415.3 | | |
| | | 4 | 0.2804 | 1845.82+0.2(10206.8)+0.65(10927.6)+0.15(10842.7)=12616.5 | | |
| | | 5 | 0.3199 | 313.76+0.15(10206.8)+0.75(10927.6)+0.1(10842.7)=11124.8 | | |
| 9 | 240.35 | 1 | 0.173 | 9206.44+0.75(10927.6)+0.25(10842.7)=20112.8 | 21825.5 | 2 |
| | | 2 | 0.2046 | 10927.64+0.65(10927.6)+0.35(10842.7)=21825.5 | | |
| | | 3 | 0.2442 | 7327.75+0.5(10927.6)+0.5(10842.7)=18212.9 | | |
| | | 4 | 0.2869 | 2556.97+0.35(10927.6)+0.65(10842.7)=13429.4 | | |
| | | 5 | 0.3308 | 451.9+0.25(10927.6)+0.75(10842.7)=11315.8 | | |
| 10 | 262.94 | 1 | 0.1835 | 9701.99+0.2(10927.6)+0.8(10842.7)=20561.6 | 21698 | 2 |
| | | 2 | 0.2203 | 10842.65+0.15(10927.6)+0.85(10842.7) =21698 | | |
| | | 3 | 0.2648 | 6355.54+0.1(10927.6)+0.9(10842.7)=17206.7 | | |
| | | 4 | 0.3122 | 1810.01+0.05(10927.6)+0.95(10842.7)=12656.9 | | |
| | | 5 | 0.3608 | 244.43+(10842.7)=11087.1 | | |

The calculations for 7th hole are shown in Table 4.

$$v_i(7) = \max_k [q_i^k + p_{i1}^k v_1(8) + p_{i2}^k v_2(8) + p_{i3}^k v_3(8) + p_{i4}^k v_4(8) + p_{i5}^k v_5(8) + p_{i6}^k v_6(8) + p_{i7}^k v_7(8)$$
$$+ p_{i8}^k v_8(8) + p_{i9}^k v_9(8) + p_{i10}^k v_{10}(8)]$$
$$= \max_k [q_i^k + p_{i1}^k (15377.3) + p_{i2}^k (17300) + p_{i3}^k (18968.8) + p_{i4}^k (18719.7)$$
$$+ p_{i5}^k (20448.5) + p_{i6}^k (20566.5) + p_{i7}^k (20922.9) + p_{i8}^k (20613.1)$$
$$+ p_{i9}^k v_9 (21825.5) + p_{i10}^k (21698)]$$

At the end of hole 7, where n=7, the optimal decision is to select the second alternative in each state. Consequently, the decision vector is computed as d(7)=[2 2 2 2 2 2 2 2 2]$^T$.

TABLE IV
DATA VALUE ITERATION FOR N=7

| State i | | Decision k | | $v_i(7)$ | Expected total reward, x10$^{-2}$ mm | Decision |
|---|---|---|---|---|---|---|
| 1 | 50.92 | 1 | 0.0522 | 23186.7 | 23730.7 | 2 |
| | | 2 | 0.0582 | 23730.7 | | |
| | | 3 | 0.0658 | 21347.9 | | |
| | | 4 | 0.0742 | 19222.1 | | |
| | | 5 | 0.0831 | 17676.6 | | |
| 2 | 83.75 | 1 | 0.0832 | 25970.5 | 26347.1 | 2 |
| | | 2 | 0.0931 | 26347.1 | | |
| | | 3 | 0.1053 | 24171.3 | | |
|   |        | 4 | 0.1185 | 21425.6 |         |   |
|   |        | 5 | 0.1324 | 19548   |         |   |
| 3 | 112.12 | 1 | 0.1029 | 26911.4 | 28591.5 | 2 |
|   |        | 2 | 0.1162 | 28591.5 |         |   |
|   |        | 3 | 0.1341 | 26316   |         |   |
|   |        | 4 | 0.154  | 22597.7 |         |   |
|   |        | 5 | 0.1749 | 19925.8 |         |   |
| 4 | 147.36 | 1 | 0.1254 | 28392.2 | 28483.5 | 2 |
|   |        | 2 | 0.1481 | 28483.5 |         |   |
|   |        | 3 | 0.1753 | 24315.1 |         |   |
|   |        | 4 | 0.2043 | 21312.9 |         |   |
|   |        | 5 | 0.2341 | 20343.7 |         |   |
| 5 | 159.54 | 1 | 0.1316 | 28574.9 | 30741.4 | 2 |
|   |        | 2 | 0.1513 | 30741.4 |         |   |
|   |        | 3 | 0.1781 | 28366.2 |         |   |
|   |        | 4 | 0.2077 | 23940.6 |         |   |
|   |        | 5 | 0.2387 | 21342.9 |         |   |
| 6 | 187.06 | 1 | 0.1461 | 29534.2 | 30914.8 | 2 |
|   |        | 2 | 0.1729 | 30914.8 |         |   |
|   |        | 3 | 0.2067 | 27035.2 |         |   |
|   |        | 4 | 0.2431 | 22676.3 |         |   |
|   |        | 5 | 0.2807 | 21100.8 |         |   |
| 7 | 203.92 | 1 | 0.1602 | 30589.3 | 31446.2 | 2 |
|   |        | 2 | 0.1863 | 31446.2 |         |   |
|   |        | 3 | 0.2177 | 27858.6 |         |   |
|   |        | 4 | 0.2512 | 23698.5 |         |   |
|   |        | 5 | 0.2858 | 21439.2 |         |   |
| 8 | 228.71 | 1 | 0.1738 | 31110.2 | 31185.1 | 2 |
|   |        | 2 | 0.2053 | 31185.1 |         |   |
|   |        | 3 | 0.2418 | 27067.4 |         |   |
|   |        | 4 | 0.2804 | 23409.7 |         |   |
|   |        | 5 | 0.3199 | 21944.7 |         |   |
| 9 | 240.35 | 1 | 0.173  | 31000.1 | 32708.6 | 2 |
|   |        | 2 | 0.2046 | 32708.6 |         |   |
|   |        | 3 | 0.2442 | 29089.5 |         |   |
|   |        | 4 | 0.2869 | 24299.6 |         |   |
|   |        | 5 | 0.3308 | 22181.8 |         |   |
| 10| 262.94 | 1 | 0.1835 | 31425.5 | 32559.8 | 2 |
|   |        | 2 | 0.2203 | 32559.8 |         |   |
|   |        | 3 | 0.2648 | 28066.3 |         |   |
|   |        | 4 | 0.3122 | 23514.4 |         |   |

| | | 5 | 0.3608 | 21942.5 | | |

The calculations for 6th hole are shown in Table 5.

$$v_i(6) = \max_k[q_i^k + p_{i1}^k v_1(7) + p_{i2}^k v_2(7) + p_{i3}^k v_3(7) + p_{i4}^k v_4(7) + p_{i5}^k v_5(7) + p_{i6}^k v_6(7) + p_{i7}^k v_7(7)$$
$$+ p_{i8}^k v_8(7) + p_{i9}^k v_9(7) + p_{i10}^k v_{10}(7)]$$
$$= \max_k[q_i^k + p_{i1}^k(23730.7) + p_{i2}^k(26347.1) + p_{i3}^k(28591.5) + p_{i4}^k(28483.5)$$
$$+ p_{i5}^k(30741.4) + p_{i6}^k(30914.8) + p_{i7}^k(31446.2) + p_{i8}^k(31185.1)$$
$$+ p_{i9}^k v_9(32708.6) + p_{i10}^k(32559.8)]$$

At the end of hole 6, where n=6, the optimal decision is to select the second alternative in each state. Consequently, the decision vector is computed as d(6)=[2 2 2 2 2 2 2 2 2]$^T$.

TABLE V

DATA VALUE ITERATION FOR N=6

| State i | | Decision k | $v_i(6)$ | Expected total reward, x10$^{-2}$ mm | Decision |
|---|---|---|---|---|---|
| 1 | 50.92 | 1 | 0.0522 | 31771 | 32413.2 | 2 |
| | | 2 | 0.0582 | 32413.2 | | |
| | | 3 | 0.0658 | 30048.1 | | |
| | | 4 | 0.0742 | 28216.8 | | |
| | | 5 | 0.0831 | 26677.2 | | |
| 2 | 83.75 | 1 | 0.0832 | 35175.7 | 35616.9 | 2 |
| | | 2 | 0.0931 | 35616.9 | | |
| | | 3 | 0.1053 | 33506.2 | | |
| | | 4 | 0.1185 | 30954.4 | | |
| | | 5 | 0.1324 | 29098.5 | | |
| 3 | 112.12 | 1 | 0.1029 | 36622.3 | 38342.9 | 2 |
| | | 2 | 0.1162 | 38342.9 | | |
| | | 3 | 0.1341 | 36009.2 | | |
| | | 4 | 0.154 | 32412.7 | | |
| | | 5 | 0.1749 | 29721.4 | | |
| 4 | 147.36 | 1 | 0.1254 | 38293.8 | 38440.9 | 2 |
| | | 2 | 0.1481 | 38440.9 | | |
| | | 3 | 0.1753 | 34343.5 | | |
| | | 4 | 0.2043 | 31508.3 | | |
| | | 5 | 0.2341 | 30562.8 | | |
| 5 | 159.54 | 1 | 0.1316 | 38899.1 | 41080 | 2 |
| | | 2 | 0.1513 | 41080 | | |
| | | 3 | 0.1781 | 38686.8 | | |
| | | 4 | 0.2077 | 34304 | | |
| | | 5 | 0.2387 | 31700.4 | | |
| 6 | 187.06 | 1 | 0.1461 | 39931.1 | 41331.6 | 2 |

| | | 2 | 0.1729 | 41331.6 | | |
|---|---|---|---|---|---|---|
| | | 3 | 0.2067 | 37471 | | |
| | | 4 | 0.2431 | 33171.9 | | |
| | | 5 | 0.2807 | 31602.8 | | |
| 7 | 203.92 | 1 | 0.1602 | 41155.9 | 42033.3 | 2 |
| | | 2 | 0.1863 | 42033.3 | | |
| | | 3 | 0.2177 | 38406.3 | | |
| | | 4 | 0.2512 | 34307.4 | | |
| | | 5 | 0.2858 | 32035 | | |
| 8 | 228.71 | 1 | 0.1738 | 41757.8 | 41862.8 | 2 |
| | | 2 | 0.2053 | 41862.8 | | |
| | | 3 | 0.2418 | 37795 | | |
| | | 4 | 0.2804 | 34227.4 | | |
| | | 5 | 0.3199 | 32778.9 | | |
| 9 | 240.35 | 1 | 0.173 | 41877.8 | 43584.1 | 2 |
| | | 2 | 0.2046 | 43584.1 | | |
| | | 3 | 0.2442 | 39961.9 | | |
| | | 4 | 0.2869 | 35168.8 | | |
| | | 5 | 0.3308 | 33048.9 | | |
| 10 | 262.94 | 1 | 0.1835 | 42291.6 | 43424.8 | 2 |
| | | 2 | 0.2203 | 43424.8 | | |
| | | 3 | 0.2648 | 38930.2 | | |
| | | 4 | 0.3122 | 34377.3 | | |
| | | 5 | 0.3608 | 32804.3 | | |

The calculations for 5th hole are shown in Table 6.

$$v_i(5) = \max_k [q_i^k + p_{i1}^k v_1(6) + p_{i2}^k v_2(6) + p_{i3}^k v_3(6) + p_{i4}^k v_4(6) + p_{i5}^k v_5(6) + p_{i6}^k v_6(6) + p_{i7}^k v_7(6)$$
$$+ p_{i8}^k v_8(6) + p_{i9}^k v_9(6) + p_{i10}^k v_{10}(6)]$$
$$= \max_k [q_i^k + p_{i1}^k(32413.2) + p_{i2}^k(35616.9) + p_{i3}^k(38342.9) + p_{i4}^k(38440.9)$$
$$+ p_{i5}^k(41080) + p_{i6}^k(41331.6) + p_{i7}^k(42033.3) + p_{i8}^k(41862.8) + p_{i9}^k v_9(43584.1)$$
$$+ p_{i10}^k(43424.8)]$$

At the end of hole 5, where n=5, the optimal decision is to select the second alternative in each state. Consequently, the decision vector is computed as d(5)=[2 2 2 2 2 2 2 2 2 2]$^T$.

TABLE VI
DATA VALUE ITERATION FOR N=5

| State i | | Decision k | $v_i(6)$ | Expected total reward, x10$^{-2}$ mm | Decision |
|---|---|---|---|---|---|
| 1 | 50.92 | 1 | 0.0522 | 40648.5 | 41373.5 | 2 |
| | | 2 | 0.0582 | 41373.5 | | |
| | | 3 | 0.0658 | 39024.3 | | |

|   |        | 4 | 0.0742 | 37441.3 |         |   |
|   |        | 5 | 0.0831 | 35907.1 |         |   |
| 2 | 83.75  | 1 | 0.0832 | 44586.4 | 45086.1 | 2 |
|   |        | 2 | 0.0931 | 45086.1 |         |   |
|   |        | 3 | 0.1053 | 43016.8 |         |   |
|   |        | 4 | 0.1185 | 40640.3 |         |   |
|   |        | 5 | 0.1324 | 38798.3 |         |   |
| 3 | 112.12 | 1 | 0.1029 | 46463.3 | 48223.6 | 2 |
|   |        | 2 | 0.1162 | 48223.6 |         |   |
|   |        | 3 | 0.1341 | 45863.6 |         |   |
|   |        | 4 | 0.154  | 42386   |         |   |
|   |        | 5 | 0.1749 | 39686   |         |   |
| 4 | 147.36 | 1 | 0.1254 | 48354.3 | 48543.4 | 2 |
|   |        | 2 | 0.1481 | 48543.4 |         |   |
|   |        | 3 | 0.1753 | 44491.5 |         |   |
|   |        | 4 | 0.2043 | 41782.4 |         |   |
|   |        | 5 | 0.2341 | 40852   |         |   |
| 5 | 159.54 | 1 | 0.1316 | 49274.3 | 51471.5 | 2 |
|   |        | 2 | 0.1513 | 51471.5 |         |   |
|   |        | 3 | 0.1781 | 49064.5 |         |   |
|   |        | 4 | 0.2077 | 44730.8 |         |   |
|   |        | 5 | 0.2387 | 42122.5 |         |   |
| 6 | 187.06 | 1 | 0.1461 | 50399.6 | 51821.6 | 2 |
|   |        | 2 | 0.1729 | 51821.6 |         |   |
|   |        | 3 | 0.2067 | 47973   |         |   |
|   |        | 4 | 0.2431 | 43738.5 |         |   |
|   |        | 5 | 0.2807 | 42173.3 |         |   |
| 7 | 203.92 | 1 | 0.1602 | 51785.4 | 52681.7 | 2 |
|   |        | 2 | 0.1863 | 52681.7 |         |   |
|   |        | 3 | 0.2177 | 49038.6 |         |   |
|   |        | 4 | 0.2512 | 44996.7 |         |   |
|   |        | 5 | 0.2858 | 42718.9 |         |   |
| 8 | 228.71 | 1 | 0.1738 | 52483.9 | 52608.1 | 2 |
|   |        | 2 | 0.2053 | 52608.1 |         |   |
|   |        | 3 | 0.2418 | 48571.6 |         |   |
|   |        | 4 | 0.2804 | 45061.8 |         |   |
|   |        | 5 | 0.3199 | 43623.8 |         |   |
| 9 | 240.35 | 1 | 0.173  | 52750.7 | 54456   | 2 |
|   |        | 2 | 0.2046 | 54456   |         |   |
|   |        | 3 | 0.2442 | 50832.2 |         |   |
|   |        | 4 | 0.2869 | 46037.5 |         |   |

| | | 5 | 0.3308 | 43916.5 | | |
|---|---|---|---|---|---|---|
| 10 | 262.94 | 1 | 0.1835 | 53158.6 | 54291.3 | 2 |
| | | 2 | 0.2203 | 54291.3 | | |
| | | 3 | 0.2648 | 49796.3 | | |
| | | 4 | 0.3122 | 45242.8 | | |
| | | 5 | 0.3608 | 43669.2 | | |

The calculations for 4th hole are shown in Table 7.

$$v_i(4) = \max_k [q_i^k + p_{i1}^k v_1(5) + p_{i2}^k v_2(5) + p_{i3}^k v_3(5) + p_{i4}^k v_4(5) + p_{i5}^k v_5(5) + p_{i6}^k v_6(5) + p_{i7}^k v_7(5)$$
$$+ p_{i8}^k v_8(5) + p_{i9}^k v_9(5) + p_{i10}^k v_{10}(5)]$$
$$= \max_k [q_i^k + p_{i1}^k (41373.5) + p_{i2}^k (45086.1) + p_{i3}^k (48223.6) + p_{i4}^k (48543.4)$$
$$+ p_{i5}^k (51471.5) + p_{i6}^k (51821.6) + p_{i7}^k (52681.7) + p_{i8}^k (52608.1) + p_{i9}^k v_9(54456)$$
$$+ p_{i10}^k (54291.3)]$$

At the end of hole 4, where n=4, the optimal decision is to select the second alternative in each state. Consequently, the decision vector is computed as d(4)=[2 2 2 2 2 2 2 2 2 2]$^T$.

TABLE VII
DATA VALUE ITERATION FOR N=4

| State i | | Decision k | $v_i(6)$ | Expected total reward, x10$^{-2}$ mm | Decision |
|---|---|---|---|---|---|
| 1 | 50.92 | 1 | 0.0522 | 49777.2 | 50573.6 | 2 |
| | | 2 | 0.0582 | 50573.6 | | |
| | | 3 | 0.0658 | 48239 | | |
| | | 4 | 0.0742 | 46870.5 | | |
| | | 5 | 0.0831 | 45341.1 | | |
| 2 | 83.75 | 1 | 0.0832 | 54180.7 | 54732.6 | 2 |
| | | 2 | 0.0931 | 54732.6 | | |
| | | 3 | 0.1053 | 52691.7 | | |
| | | 4 | 0.1185 | 50472 | | |
| | | 5 | 0.1324 | 48639.4 | | |
| 3 | 112.12 | 1 | 0.1029 | 56428.4 | 58225.3 | 2 |
| | | 2 | 0.1162 | 58225.3 | | |
| | | 3 | 0.1341 | 55855.2 | | |
| | | 4 | 0.154 | 52487.5 | | |
| | | 5 | 0.1749 | 49784.1 | | |
| 4 | 147.36 | 1 | 0.1254 | 58538.9 | 58761.8 | 2 |
| | | 2 | 0.1481 | 58761.8 | | |
| | | 3 | 0.1753 | 54738.5 | | |
| | | 4 | 0.2043 | 52130.8 | | |
| | | 5 | 0.2341 | 51210 | | |
| 5 | 159.54 | 1 | 0.1316 | 59706.3 | 61921.2 | 2 |

| | | 2 | 0.1513 | 61921.2 | | |
| | | 3 | 0.1781 | 59505.3 | | |
| | | 4 | 0.2077 | 55224.8 | | |
| | | 5 | 0.2387 | 52613.6 | | |
| 6 | 187.06 | 1 | 0.1461 | 60938.9 | 62381.6 | 2 |
| | | 2 | 0.1729 | 62381.6 | | |
| | | 3 | 0.2067 | 58542.2 | | |
| | | 4 | 0.2431 | 54369.8 | | |
| | | 5 | 0.2807 | 52807.7 | | |
| 7 | 203.92 | 1 | 0.1602 | 62470.7 | 63383.1 | 2 |
| | | 2 | 0.1863 | 63383.1 | | |
| | | 3 | 0.2177 | 59735.5 | | |
| | | 4 | 0.2512 | 55741.6 | | |
| | | 5 | 0.2858 | 53462.4 | | |
| 8 | 228.71 | 1 | 0.1738 | 63260.4 | 63396.9 | 2 |
| | | 2 | 0.2053 | 63396.9 | | |
| | | 3 | 0.2418 | 59380.2 | | |
| | | 4 | 0.2804 | 55907.5 | | |
| | | 5 | 0.3199 | 54476.1 | | |
| 9 | 240.35 | 1 | 0.173 | 63621.3 | 65326 | 2 |
| | | 2 | 0.2046 | 65326 | | |
| | | 3 | 0.2442 | 61701.4 | | |
| | | 4 | 0.2869 | 56905.9 | | |
| | | 5 | 0.3308 | 54784.4 | | |
| 10 | 262.94 | 1 | 0.1835 | 64026.3 | 65158.7 | 2 |
| | | 2 | 0.2203 | 65158.7 | | |
| | | 3 | 0.2648 | 60663.3 | | |
| | | 4 | 0.3122 | 56109.6 | | |
| | | 5 | 0.3608 | 54535.8 | | |

The calculations for 3rd hole are shown in Table 8.

$$v_i(3) = \max_k [q_i^k + p_{i1}^k v_1(4) + p_{i2}^k v_2(4) + p_{i3}^k v_3(4) + p_{i4}^k v_4(4) + p_{i5}^k v_5(4) + p_{i6}^k v_6(4) + p_{i7}^k v_7(4)$$
$$+ p_{i8}^k v_8(4) + p_{i9}^k v_9(4) + p_{i10}^k v_{10}(4)]$$
$$= \max_k [q_i^k + p_{i1}^k (50573.6) + p_{i2}^k (54732.6) + p_{i3}^k (58225.3) + p_{i4}^k (58761.8)$$
$$+ p_{i5}^k (61921.2) + p_{i6}^k (62381.6) + p_{i7}^k (63383.1) + p_{i8}^k (63396.9) + p_{i9}^k v_9 (65326)$$
$$+ p_{i10}^k (65158.7)]$$

At the end of hole 3, where n=3, the optimal decision is to select the second alternative in each state. Consequently, the decision vector is computed as d(3)=[2 2 2 2 2 2 2 2 2 2]$^T$.

TABLE VIII
DATA VALUE ITERATION FOR N=3

| State i | | Decision k | $v_i(6)$ | Expected total reward, x10⁻² mm | Decision |
|---|---|---|---|---|---|
| 1 | 50.92 | 1 | 0.0522 | 59124.4 | 59983.2 | 2 |
| | | 2 | 0.0582 | 59983.2 | | |
| | | 3 | 0.0658 | 57662.3 | | |
| | | 4 | 0.0742 | 56481 | | |
| | | 5 | 0.0831 | 54956.1 | | |
| 2 | 83.75 | 1 | 0.0832 | 63937.6 | 64535.9 | 2 |
| | | 2 | 0.0931 | 64535.9 | | |
| | | 3 | 0.1053 | 62515.8 | | |
| | | 4 | 0.1185 | 60435.1 | | |
| | | 5 | 0.1324 | 58609.5 | | |
| 3 | 112.12 | 1 | 0.1029 | 66507.3 | 68337.5 | 2 |
| | | 2 | 0.1162 | 68337.5 | | |
| | | 3 | 0.1341 | 65965.2 | | |
| | | 4 | 0.154 | 62697.3 | | |
| | | 5 | 0.1749 | 59993.1 | | |
| 4 | 147.36 | 1 | 0.1254 | 68826.2 | 69077.7 | 2 |
| | | 2 | 0.1481 | 69077.7 | | |
| | | 3 | 0.1753 | 65072.5 | | |
| | | 4 | 0.2043 | 62550.9 | | |
| | | 5 | 0.2341 | 61636 | | |
| 5 | 159.54 | 1 | 0.1316 | 70197.7 | 72430.7 | 2 |
| | | 2 | 0.1513 | 72430.7 | | |
| | | 3 | 0.1781 | 70010.2 | | |
| | | 4 | 0.2077 | 65784 | | |
| | | 5 | 0.2387 | 63171.2 | | |
| 6 | 187.06 | 1 | 0.1461 | 71543 | 73004.2 | 2 |
| | | 2 | 0.1729 | 73004.2 | | |
| | | 3 | 0.2067 | 69172.9 | | |
| | | 4 | 0.2431 | 65056 | | |
| | | 5 | 0.2807 | 63496.6 | | |
| 7 | 203.92 | 1 | 0.1602 | 73202 | 74127.2 | 2 |
| | | 2 | 0.1863 | 74127.2 | | |
| | | 3 | 0.2177 | 70480.6 | | |
| | | 4 | 0.2512 | 66525.1 | | |
| | | 5 | 0.2858 | 64246.2 | | |
| 8 | 228.71 | 1 | 0.1738 | 74069.2 | 74213.8 | 2 |
| | | 2 | 0.2053 | 74213.8 | | |
| | | 3 | 0.2418 | 70209.6 | | |
| | | 4 | 0.2804 | 66760.9 | | |
| | | 5 | 0.3199 | 65333.7 | | |

| | | 1 | 0.173 | 74490.6 | | |
|---|---|---|---|---|---|---|
| | | 2 | 0.2046 | 76195.1 | | |
| 9 | 240.35 | 3 | 0.2442 | 72570.1 | 76195.1 | 2 |
| | | 4 | 0.2869 | 67774.2 | | |
| | | 5 | 0.3308 | 65652.4 | | |
| | | 1 | 0.1835 | 74894.1 | | |
| | | 2 | 0.2203 | 76026.4 | | |
| 10 | 262.94 | 3 | 0.2648 | 71531 | 76026.4 | 2 |
| | | 4 | 0.3122 | 66977.1 | | |
| | | 5 | 0.3608 | 65403.1 | | |

The calculations for 2nd hole are shown in Table 9.

$$v_i(2) = \max_k [q_i^k + p_{i1}^k v_1(3) + p_{i2}^k v_2(3) + p_{i3}^k v_3(3) + p_{i4}^k v_4(3) + p_{i5}^k v_5(3) + p_{i6}^k v_6(3) + p_{i7}^k v_7(3)$$
$$+ p_{i8}^k v_8(3) + p_{i9}^k v_9(3) + p_{i10}^k v_{10}(3)]$$
$$= \max_k [q_i^k + p_{i1}^k (59983.2) + p_{i2}^k (64535.9) + p_{i3}^k (68337.5) + p_{i4}^k (69077.7)$$
$$+ p_{i5}^k (72430.7) + p_{i6}^k (73004.2) + p_{i7}^k (74127.2) + p_{i8}^k (74213.8)$$
$$+ p_{i9}^k (76195.1) + p_{i10}^k (76026.4)]$$

At the end of hole 2, where n=2, the optimal decision is to select the second alternative in each state. Consequently, the decision vector is computed as d(2)=[2 2 2 2 2 2 2 2 2 2]$^T$.

TABLE IX
DATA VALUE ITERATION FOR N=2

| State i | | Decision k | $v_i(6)$ | Expected total reward, x10$^{-2}$ mm | Decision |
|---|---|---|---|---|---|
| | | 1 | 0.0522 | 68663.4 | | |
| | | 2 | 0.0582 | 69577 | | |
| 1 | 50.92 | 3 | 0.0658 | 67268.8 | 69577 | 2 |
| | | 4 | 0.0742 | 66251.9 | | |
| | | 5 | 0.0831 | 64731.3 | | |
| | | 1 | 0.0832 | 73838.5 | | |
| | | 2 | 0.0931 | 74477.9 | | |
| 2 | 83.75 | 3 | 0.1053 | 72473.6 | 74477.9 | 2 |
| | | 4 | 0.1185 | 70516.1 | | |
| | | 5 | 0.1324 | 68695.7 | | |
| | | 1 | 0.1029 | 76689.9 | | |
| | | 2 | 0.1162 | 78550.1 | | |
| 3 | 112.12 | 3 | 0.1341 | 76179.3 | 78550.1 | 2 |
| | | 4 | 0.154 | 73001.5 | | |
| | | 5 | 0.1749 | 70297.9 | | |
| 4 | 147.36 | 1 | 0.1254 | 79201.8 | 79478.3 | 2 |
| | | 2 | 0.1481 | 79478.3 | | |

| | | 3 | 0.1753 | 75485.3 | | |
| | | 4 | 0.2043 | 73038.6 | | |
| | | 5 | 0.2341 | 72127.8 | | |
| 5 | 159.54 | 1 | 0.1316 | 80747.7 | 82998.1 | 2 |
| | | 2 | 0.1513 | 82998.1 | | |
| | | 3 | 0.1781 | 80576.2 | | |
| | | 4 | 0.2077 | 76402.2 | | |
| | | 5 | 0.2387 | 73789 | | |
| 6 | 187.06 | 1 | 0.1461 | 82203.2 | 83680.3 | 2 |
| | | 2 | 0.1729 | 83680.3 | | |
| | | 3 | 0.2067 | 79856.2 | | |
| | | 4 | 0.2431 | 75786.8 | | |
| | | 5 | 0.2807 | 74229.8 | | |
| 7 | 203.92 | 1 | 0.1602 | 83969.6 | 84904.6 | 2 |
| | | 2 | 0.1863 | 84904.6 | | |
| | | 3 | 0.2177 | 81261.1 | | |
| | | 4 | 0.2512 | 77335.2 | | |
| | | 5 | 0.2858 | 75057.3 | | |
| 8 | 228.71 | 1 | 0.1738 | 84899 | 85048.7 | 2 |
| | | 2 | 0.2053 | 85048.7 | | |
| | | 3 | 0.2418 | 81052.6 | | |
| | | 4 | 0.2804 | 77619.3 | | |
| | | 5 | 0.3199 | 76194.8 | | |
| 9 | 240.35 | 1 | 0.173 | 85359.4 | 87063.7 | 2 |
| | | 2 | 0.2046 | 87063.7 | | |
| | | 3 | 0.2442 | 83438.5 | | |
| | | 4 | 0.2869 | 78642.4 | | |
| | | 5 | 0.3308 | 76520.5 | | |
| 10 | 262.94 | 1 | 0.1835 | 85762.2 | 86894.4 | 2 |
| | | 2 | 0.2203 | 86894.4 | | |
| | | 3 | 0.2648 | 82398.8 | | |
| | | 4 | 0.3122 | 77844.9 | | |
| | | 5 | 0.3608 | 76270.9 | | |

The calculations for 1st hole are shown in Table 10.

$$v_i(1) = \max_k [q_i^k + p_{i1}^k v_1(2) + p_{i2}^k v_2(2) + p_{i3}^k v_3(2) + p_{i4}^k v_4(2) + p_{i5}^k v_5(2) + p_{i6}^k v_6(2) + p_{i7}^k v_7(2)$$
$$+ p_{i8}^k v_8(2) + p_{i9}^k v_9(2) + p_{i10}^k v_{10}(2)]$$
$$= \max_k [q_i^k + p_{i1}^k (69577) + p_{i2}^k (74477.9) + p_{i3}^k (78550.1) + p_{i4}^k (79478.3)$$
$$+ p_{i5}^k (82998.1) + p_{i6}^k (83680.3) + p_{i7}^k (84904.6) + p_{i8}^k (85048.7)$$
$$+ p_{i9}^k v_9(87063.7) + p_{i10}^k (86894.4)]$$

At the end of hole 1, where n=1, the optimal decision is to select the second alternative in each state. Consequently, the decision vector is computed as d(1)=[2 2 2 2 2 2 2 2 2]$^T$.

TABLE X
DATA VALUE ITERATION FOR N=1

| State i | | Decision k | $v_i(6)$ | Expected total reward, x10$^{-2}$ mm | Decision |
|---|---|---|---|---|---|
| 1 | 50.92 | 1 | 0.0522 | 78371.3 | |
| | | 2 | 0.0582 | 79333.2 | |
| | | 3 | 0.0658 | 77036.7 | 79333.2 | 2 |
| | | 4 | 0.0742 | 76164.8 | |
| | | 5 | 0.0831 | 74648.1 | |
| 2 | 83.75 | 1 | 0.0832 | 83867 | |
| | | 2 | 0.0931 | 84542.8 | |
| | | 3 | 0.1053 | 82550.8 | 84542.8 | 2 |
| | | 4 | 0.1185 | 80702.8 | |
| | | 5 | 0.1324 | 78886.5 | |
| 3 | 112.12 | 1 | 0.1029 | 86966.1 | |
| | | 2 | 0.1162 | 88853.5 | |
| | | 3 | 0.1341 | 86485.9 | 88853.5 | 2 |
| | | 4 | 0.154 | 83389.5 | |
| | | 5 | 0.1749 | 80687 | |
| 4 | 147.36 | 1 | 0.1254 | 89655 | |
| | | 2 | 0.1481 | 89953.6 | |
| | | 3 | 0.1753 | 85969.3 | 89953.6 | 2 |
| | | 4 | 0.2043 | 83588.9 | |
| | | 5 | 0.2341 | 82681 | |
| 5 | 159.54 | 1 | 0.1316 | 91352.3 | |
| | | 2 | 0.1513 | 93618.7 | |
| | | 3 | 0.1781 | 91197.9 | 93618.7 | 2 |
| | | 4 | 0.2077 | 87071.7 | |
| | | 5 | 0.2387 | 84458.9 | |
| 6 | 187.06 | 1 | 0.1461 | 92910.4 | |
| | | 2 | 0.1729 | 94400.4 | |
| | | 3 | 0.2067 | 90583 | 94400.4 | 2 |
| | | 4 | 0.2431 | 86552.5 | |
| | | 5 | 0.2807 | 84997.7 | |
| 7 | 203.92 | 1 | 0.1602 | 94764.8 | |
| | | 2 | 0.1863 | 95707.2 | |
| | | 3 | 0.2177 | 92067.3 | 95707.2 | 2 |
| | | 4 | 0.2512 | 88163.7 | |
| | | 5 | 0.2858 | 85887 | |
| 8 | 228.71 | 1 | 0.1738 | 95742.3 | 95895.3 | 2 |

| | | 2 | 0.2053 | 95895.3 | | |
|---|---|---|---|---|---|---|
| | | 3 | 0.2418 | 91904.3 | | |
| | | 4 | 0.2804 | 88481.1 | | |
| | | 5 | 0.3199 | 87058.3 | | |
| 9 | 240.35 | 1 | 0.173 | 96227.8 | 97932.1 | 2 |
| | | 2 | 0.2046 | 97932.1 | | |
| | | 3 | 0.2442 | 94306.8 | | |
| | | 4 | 0.2869 | 89510.6 | | |
| | | 5 | 0.3308 | 87388.6 | | |
| 10 | 262.94 | 1 | 0.1835 | 96630.2 | 97762.4 | 2 |
| | | 2 | 0.2203 | 97762.4 | | |
| | | 3 | 0.2648 | 93266.9 | | |
| | | 4 | 0.3122 | 88712.9 | | |
| | | 5 | 0.3608 | 87138.8 | | |

Finally, the calculations for hole 0 are indicated in Table 11.

$$v_i(0) = \max_k [q_i^k + p_{i1}^k v_1(1) + p_{i2}^k v_2(1) + p_{i3}^k v_3(1) + p_{i4}^k v_4(1) + p_{i5}^k v_5(1) + p_{i6}^k v_6(1) + p_{i7}^k v_7(1)$$
$$+ p_{i8}^k v_8(1) + p_{i9}^k v_9(1) + p_{i10}^k v_{10}(1)]$$
$$= \max_k [q_i^k + p_{i1}^k (79333.2) + p_{i2}^k (84542.8) + p_{i3}^k (88853.5) + p_{i4}^k (89953.6)$$
$$+ p_{i5}^k (93618.7) + p_{i6}^k (94400.4) + p_{i7}^k (95707.2) + p_{i8}^k (95895.3)$$
$$+ p_{i9}^k v_9 (97932.1) + p_{i10}^k (97762.4)]$$

At the end of hole 0, which is the beginning of hole 1, the optimal decision is to select the second alternative in each state. Consequently, the decision vector is computed as d(0)=[2 2 2 2 2 2 2 2 2]$^T$.

TABLE XI
DATA VALUE ITERATION FOR N=0

| State i | | Decision k | $v_i(6)$ | Expected total reward, x10$^{-2}$ mm | Decision |
|---|---|---|---|---|---|
| 1 | 50.92 | 1 | 0.0522 | 88228.5 | 89233.3 | 2 |
| | | 2 | 0.0582 | 89233.3 | | |
| | | 3 | 0.0658 | 86203.8 | | |
| | | 4 | 0.0742 | 86203.8 | | |
| | | 5 | 0.0831 | 84690.6 | | |
| 2 | 83.75 | 1 | 0.0832 | 94008.7 | 94716.9 | 2 |
| | | 2 | 0.0931 | 94716.9 | | |
| | | 3 | 0.1053 | 92735 | | |
| | | 4 | 0.1185 | 90984.3 | | |
| | | 5 | 0.1324 | 89171.3 | | |
| 3 | 112.12 | 1 | 0.1029 | 97327 | 99238.8 | 2 |
| | | 2 | 0.1162 | 99238.8 | | |
| | | 3 | 0.1341 | 96875.3 | | |

|   |        | 4 | 0.154  | 93852.3 |        |   |
|   |        | 5 | 0.1749 | 91151   |        |   |
| 4 | 147.36 | 1 | 0.1254 | 100177  | 100495 | 2 |
|   |        | 2 | 0.1481 | 100495  |        |   |
|   |        | 3 | 0.1753 | 96517.2 |        |   |
|   |        | 4 | 0.2043 | 94195.4 |        |   |
|   |        | 5 | 0.2341 | 93289.8 |        |   |
| 5 | 159.54 | 1 | 0.1316 | 102006  | 104286 | 2 |
|   |        | 2 | 0.1513 | 104286  |        |   |
|   |        | 3 | 0.1781 | 101868  |        |   |
|   |        | 4 | 0.2077 | 97784.3 |        |   |
|   |        | 5 | 0.2387 | 95172.3 |        |   |
| 6 | 187.06 | 1 | 0.1461 | 103656  | 105156 | 2 |
|   |        | 2 | 0.1729 | 105156  |        |   |
|   |        | 3 | 0.2067 | 101344  |        |   |
|   |        | 4 | 0.2431 | 97345.2 |        |   |
|   |        | 5 | 0.2807 | 95792.4 |        |   |
| 7 | 203.92 | 1 | 0.1602 | 105581  | 106529 | 2 |
|   |        | 2 | 0.1863 | 106529  |        |   |
|   |        | 3 | 0.2177 | 102892  |        |   |
|   |        | 4 | 0.2512 | 99004.8 |        |   |
|   |        | 5 | 0.2858 | 96729.2 |        |   |
| 8 | 228.71 | 1 | 0.1738 | 106594  | 106750 | 2 |
|   |        | 2 | 0.2053 | 106750  |        |   |
|   |        | 3 | 0.2418 | 102762  |        |   |
|   |        | 4 | 0.2804 | 99345.1 |        |   |
|   |        | 5 | 0.3199 | 97923.4 |        |   |
| 9 | 240.35 | 1 | 0.173  | 107096  | 108800 | 2 |
|   |        | 2 | 0.2046 | 108800  |        |   |
|   |        | 3 | 0.2442 | 105175  |        |   |
|   |        | 4 | 0.2869 | 100379  |        |   |
|   |        | 5 | 0.3308 | 98256.7 |        |   |
| 10| 262.94 | 1 | 0.1835 | 107498  | 108631 | 2 |
|   |        | 2 | 0.2203 | 108631  |        |   |
|   |        | 3 | 0.2648 | 104135  |        |   |
|   |        | 4 | 0.3122 | 99580.9 |        |   |
|   |        | 5 | 0.3608 | 98006.9 |        |   |

### III. RESULTS AND DISCUSSION

The results of presented above calculations for the expected total rewards and the optimal decisions at the end of each hole of the 10-hole planning horizon are displayed in Table 12, 13. As it could be clearly seen from results that if the process starts at state 10, the expected total reward would be 108630.525 x10-2 mm, which is the highest for any other state. However, if the process starts at state 1, the expected total reward is 89233.2667 x10-2 mm, which would be the lowest among other states. The decision matrixes show slight change in states 4 and 8 at the hole number 9, but apart from that, decision 2 is showing dominance throughout the process.

TABLE XII

EXPECTED TOTAL REWARDS FOR f PLANNING HORIZON OF 10 HOLES

| End of hole | Expected total reward, x10$^{-2}$ mm | | | | | | | | | | |
|---|---|---|---|---|---|---|---|---|---|---|---|
| | 0 | 1 | 2 | 3 | 4 | 5 | 6 | 7 | 8 | 9 | 10 |
| $v_1(n)$ | 89233.2 | 79333.2 | 69576.9 | 59983.2 | 50573.6 | 41373.4 | 32413.1 | 23730.6 | 15377.3 | 7430.09 | 0 |
| $v_2(n)$ | 94716.9 | 84542.7 | 74477.8 | 64535.8 | 54732.5 | 45086.0 | 35616.8 | 26347.1 | 17300.0 | 8500.39 | 0 |
| $v_3(n)$ | 99238.8 | 88853.4 | 78550.0 | 68337.5 | 58225.2 | 48223.6 | 38342.9 | 28591.4 | 18968.8 | 9450.55 | 0 |
| $v_4(n)$ | 100494. | 89953.6 | 79478.3 | 69077.7 | 58761.7 | 48543.3 | 38440.8 | 28483.5 | 18719.7 | 9228.49 | 0 |
| $v_5(n)$ | 104286. | 93618.6 | 82998.0 | 72430.7 | 61921.1 | 51471.4 | 41079.9 | 30741.3 | 20448.4 | 10198.1 | 0 |
| $v_6(n)$ | 105156. | 94400.4 | 83680.2 | 73004.2 | 62381.6 | 51821.6 | 41331.6 | 30914.7 | 20566.5 | 10270.0 | 0 |
| $v_7(n)$ | 106528. | 95707.2 | 84904.6 | 74127.2 | 63383.0 | 52681.7 | 42033.2 | 31446.2 | 20922.8 | 10449.9 | 0 |
| $v_8(n)$ | 106749. | 95895.3 | 85048.7 | 74213.7 | 63396.9 | 52608.1 | 41862.7 | 31185.0 | 20613.0 | 10206.7 | 0 |
| $v_9(n)$ | 108800. | 97932.0 | 87063.6 | 76195.0 | 65326.0 | 54456.0 | 43584.1 | 32708.5 | 21825.5 | 10927.6 | 0 |
| $v_{10}(n)$ | 108630. | 97762.4 | 86894.3 | 76026.4 | 65158.6 | 54291.3 | 43424.7 | 32559.8 | 21698.0 | 10842.6 | 0 |

TABLE XIII

EXPECTED OPTIMAL DECISIONS FOR f PLANNING HORIZON OF 10 HOLES

| Decisions for each state | Decision at each state, n | | | | | | | | | | |
|---|---|---|---|---|---|---|---|---|---|---|---|
| | 0 | 1 | 2 | 3 | 4 | 5 | 6 | 7 | 8 | 9 | 10 |
| $d_1(n)$ | 2 | 2 | 2 | 2 | 2 | 2 | 2 | 2 | 2 | 2 | - |
| $d_2(n)$ | 2 | 2 | 2 | 2 | 2 | 2 | 2 | 2 | 2 | 2 | - |
| $d_3(n)$ | 2 | 2 | 2 | 2 | 2 | 2 | 2 | 2 | 2 | 2 | - |
| $d_4(n)$ | 2 | 2 | 2 | 2 | 2 | 2 | 2 | 2 | 2 | 1 | - |
| $d_5(n)$ | 2 | 2 | 2 | 2 | 2 | 2 | 2 | 2 | 2 | 2 | - |
| $d_6(n)$ | 2 | 2 | 2 | 2 | 2 | 2 | 2 | 2 | 2 | 2 | - |
| $d_7(n)$ | 2 | 2 | 2 | 2 | 2 | 2 | 2 | 2 | 2 | 2 | - |
| $d_8(n)$ | 2 | 2 | 2 | 2 | 2 | 2 | 2 | 2 | 2 | 1 | - |
| $d_9(n)$ | 2 | 2 | 2 | 2 | 2 | 2 | 2 | 2 | 2 | 2 | - |
| $d_{10}(n)$ | 2 | 2 | 2 | 2 | 2 | 2 | 2 | 2 | 2 | 2 | - |

Based on axial force value conditions the optimal decisions for feed rate could be chosen from Table 13. For each state of axial force there exists a specific optimal feed rate value, which would lead to higher reward than other feed rate values. General trend seem that the feed rate increases with the increase of axial force value. However, if all possible decisions for one state could be compared and examined, it can be seen that the optimal ones will not be the highest or the lowest values, but rather feed rate values close to the middle range. This is due to the fact that the lower feeds provides lower cutting speed, which seem to increase the total life of the tool, but appear to decrease the length of the hole being drilled for particular amount of time. In the other hand, higher feeds are likely to be more inefficient in terms of hole length because the drills operating on high speeds are subjected to excessive wear. The worn out tools could not have the same cutting speed, even when operating at higher speeds. To summarize, the optimal feed rate decisions for each axial force value during drilling deep holes are computed and presented in the given case study.

## IV. CONCLUSION

Engineers working in the manufacturing area could adopt more effective ways for feed rate policies in spiral drilling. For instance, spiral drilling requires most optimal choice of the feed rate depending upon the hardness of the material and geometry of the tool. In other words, even slight deviations in the feed rate can cause deviation of the hole straightness or even breakage of the tool inside of a hole, which will lead to the wastage of the workpiece. In addition, slight deviations in the feed rate could cause the failure of the tool, which will waste the time spent for drilling. In this paper, the feed rate optimization model based on a MDP was introduced for spiral drilling process. In particular, the experimental data on drilling was implemented for 10 states of axial force parameters with 5 feed rate decisions made in each of the states, having the length of a hole being drilled as a reward. Proposed optimization model was computed using value iteration method for 10 holes planning horizon. Furthermore, the results of computations were displayed as tables for optimal decision as well as the expected total reward in each state. In conclusion, adaptive choice of the feed rates based on MDP model is claimed to improve the efficiency of the spiral drilling in terms of cost and time.

## ACKNOWLEDGMENT

Authors would like to thank Xu Huan, Assistant Professor at National University of Singapore, for providing scientific guidance and support.

## REFERENCES

1. G.I. Smagin, Optimization of drilling conditions by minimum cost criterion (Monograph). Novosibirsk, NSTU, 1999, pp. 55-58.
2. M.L. Puterman, Markov decision processes: discrete stochastic dynamic programming. vol. 414. John Wiley & Sons, 2009.
3. S.K. Abeygunawardane, P. Jirutitijaroen, and H. Xu, "Adaptive maintenance policies for aging devices using a Markov Decision Process," IEEE Trans.on Power Systems, vol.28, no.3, 2013, pp. 3194-3203.